\documentclass[11pt]{article}

\usepackage{bm}
\usepackage{epsf,amsfonts,amsmath}
\usepackage{amsmath}
\usepackage{amssymb}
\usepackage{amsthm}

\usepackage{epic,eepic}
\usepackage{texdraw}
\usepackage[dvips]{color}
\usepackage{color}
\usepackage{graphicx}
\usepackage{exscale,relsize}
\newcommand{\field}[1]{\mathbb{#1}}
\title{\textbf{Darboux transformations, finite reduction groups and related Yang-Baxter maps}}
\author{S. Konstantinou-Rizos and A. V. Mikhailov\\
\small \textit{Department of Applied Mathematics, University of Leeds, Leeds}\\
\small \texttt{mmskr@leeds.ac.uk, A.V.Mikhailov@leeds.ac.uk}
}
\begin{document}

\maketitle
\begin{abstract}
In this paper we construct Yang-Baxter (YB) maps using Darboux matrices which are invariant under the action of finite reduction groups. We present 6-dimensional YB maps corresponding to Darboux transformations for the Nonlinear Schr\"odinger (NLS) equation and the derivative Nonlinear Schr\"odinger (DNLS) equation. These YB maps can be restricted to $4-$dimensional YB maps on invariant leaves. The former are completely integrable and they also have applications to a recent theory of maps preserving functions with symmetries \cite{Allan-Pavlos}. We give a $6-$ dimensional YB-map corresponding to the Darboux transformation for a deformation of the DNLS equation. We also consider vector generalisations of the YB maps corresponding to the NLS and DNLS equation.
\end{abstract}

\section{Introduction}
The Yang-Baxter equation has a fundamental role in the theory of quantum and classical integrable systems. In particular, Yang-Baxter maps, namely the set theoretical solutions \cite{Drinfel'd} of the Yang-Baxter equation have been of great interest for several researchers in the area of Mathematical Physics. They are related to several concepts of integrability as, for instance, the multidimensionally consistent equations \cite{ABS-2004, ABS-2005, Bobenko-Suris,Frank3,Frank5,Frank4}. Especially, for those Yang-Baxter maps which admit Lax representation \cite{Veselov2}, there are corresponding hierarchies of commuting transfer maps which preserve the spectrum of their monodromy matrix \cite{Veselov,Veselov3}. Therefore, the construction of Yang-Baxter maps is important.

There are several methods of construction of Yang-Baxter maps, coming from several fields of mathematics as the quantum group theory, representation theory, Lie group theory etc. Here, we construct Yang-Baxter maps using Darboux transformations for Lax operators of integrable partial differential equations. These Yang-Baxter maps can be reduced to completely itnegrable maps on invariant leaves.

The Yang-Baxter equation
\begin{equation}
Y^{12}\circ Y^{13} \circ Y^{23}=Y^{23}\circ Y^{13} \circ Y^{12},
\label{YB_eq1}
\end{equation}
originates in the works of Yang \cite{Yang} and Baxter \cite{Baxter}. Here $Y^{ij}$ denotes the action of a linear operator $Y:U\otimes U\rightarrow U\otimes U$ on the $ij$ factor of the triple tensor product $U \otimes U\otimes U$, where $U$ is a vector space. In this form, equation (\ref{YB_eq1}) is known in the literature as the \textit{quantum YB equation}.

Drinfel'd in 1992 \cite{Drinfel'd} proposed to replace $U$ by an arbitrary set $A$ and, therefore, the tensor product $U \otimes U$ by the Cartesian product $A \times A$. In our paper $A$ is an algebraic variety in $K^N$, where $K$ is any field of zero characteristic, such us $\field{C}$ or $\field{Q}$.

In \cite{Veselov} Veselov proposed the term \textit{Yang Baxter map} for the set-theoretical solutions of the quantum YB equation. Specifically, we consider the map $Y:A\times A\rightarrow A \times A$,
\begin{equation}
Y:(x,y)\mapsto (u(x,y),v(x,y)).
\label{Y-map}
\end{equation}
Furthermore, we define the functions $Y^{i,j}:A\times A \times A\rightarrow A\times A \times A$ for $i,j=1,2,3,~i\neq j$, which appear in equation (\ref{YB_eq1}), by the following relations
%\numparts
\begin{eqnarray}
\hspace{2.3cm} Y^{12}(x,y,z)&=&(u(x,y),v(x,y),z), \\ 
\hspace{2.3cm} Y^{13}(x,y,z)&=&(u(x,z),y,v(x,z)), \\ 
\hspace{2.3cm} Y^{23}(x,y,z)&=&(x,u(y,z),v(y,z)),
\end{eqnarray}
%\endnumparts
where $x,y,z\in A$. The variety $A$, in general, can be of any dimension. Thus, elements $x \in A$ are points in $K^N$. The map $Y^{ji}$, $i<j$, is defined as $Y^{ij}$ where we swap $u(k,l)\leftrightarrow v(l,k)$, $k,l=x,y,z$. For example, $Y^{21}(x,y,z)=(v(y,x),u(y,x),z)$.

The map (\ref{Y-map}) is a YB map, if it satisfies the YB equation ($\ref{YB_eq1}$). Moreover, it is called \textit{reversible} if the composition of $Y^{ij}$ and $Y^{ji}$ is the identity map,
\begin{equation}
Y^{ij}\circ Y^{ji}=Id.
\label{reversible}
\end{equation}

We use the term \textit{parametric YB map} when $u$ and $v$ are attached with parameters $a,b\in K^{n}$, namely $u=u(x,y;a,b)$
and $v=v(x,y;a,b)$, meaning that the following map
\begin{equation}
Y_{a,b}:(x,y;a,b)\mapsto (u(x,y;a,b),v(x,y;a,b)),
\end{equation}
satisfies the \textit{parametric YB equation}
\begin{equation}\label{YB_eq}
Y^{12}_{a,b}\circ Y^{13}_{a,c} \circ Y^{23}_{b,c}=Y^{23}_{b,c}\circ Y^{13}_{a,c} \circ Y^{12}_{a,b}.
\end{equation}

Following Suris and Veselov in \cite{Veselov2}, we call a \textit{Lax matrix for a parametric YB map}, a matrix $L=L(x;c;\lambda)$ depending on a variable $x$, a parameter $c$ and a \textit{spectral parameter} $\lambda$, such that the \textit{Lax-equation}
\begin{equation} \label{eqLax}
L(u;a,\lambda)L(v;b,\lambda)=L(y;b,\lambda)L(x;a,\lambda), \quad \text{for any $\lambda \in K$,}
\end{equation}
is satisfied due to the YB map. The above is also called a \textit{refactorisation problem}.

It is obvious that the Lax-equation (\ref{eqLax}) does not always have a unique solution, which motivated Kouloukas and Papageorgiou in \cite{Kouloukas2} to propose the term \textit{strong Lax matrix} for a YB map. This is when the Lax-equation is equivalent to a YB map
\begin{equation}\label{unique-sol}
  (u,v)=Y_{a,b}(x,y).
\end{equation}
Actually, the uniqueness of refactorisation (\ref{eqLax}) is a sufficient condition for the solutions of the Lax-equation to define a reversible YB map \cite{Kouloukas3, Kouloukas2,Veselov3} of the form (\ref{unique-sol}). In the opposite case, one may need to check if the obtained map satisfies the YB equation.

One of the most famous parametric YB maps is Adler's map \cite{Adler}
\begin{equation}\label{Adler_map}
(x_1,x_2)\longrightarrow (u,v)=\left(x_2-\frac{a-b}{x_1+x_2},x_1+\frac{a-b}{x_1+x_2}\right),
\end{equation}
which occurs from the 3-D consistent discrete potential KDV equation \cite{Frank, PNC}. In terms of Lax matrices, Adler's map (\ref{Adler_map}) is obtained from the following strong Lax matrix \cite{Veselov2, Veselov3}
\begin{equation}
L(x;a,\lambda)=
\left(\begin{matrix}
x & 1 \\
x^2+a-\lambda & x
\end{matrix}\right).
\end{equation}
In \cite{PT, PTV} a variety of YB maps is constructed using the symmetries of multi-field equations on quad graphs.

It follows from the structure of the Lax-equation (\ref{eqLax}) that we can extract \textit{invariants} of the YB map, which we denote as $I_i(x,y)$. The invariants are useful if one is interested in the dynamics of such maps. In terms of dynamics, the most interesting maps are those which are not involutive. Although, involutive maps have also useful applications \cite{Pavlos}. In all the cases presented in the next sections, our YB maps are not involutive. 

In this paper we are interested in the integrability of the YB maps as finite discrete maps. The transfer dynamics of YB maps is discussed in \cite{Veselov3}.

Now, following \cite{Fordy, Veselov4} we define integrability for YB maps.

\newtheorem{CompleteIntegrability}{Definition}[section]
\begin{CompleteIntegrability}
A $2N-$ dimensional Yang-Baxter map,  
\begin{equation}
Y:(x_1,...,x_{2N})\mapsto (u_1,...,u_{2N}), \quad u_i=u_i(x_1,...,x_{2N}), \quad i=1,...,2N, \nonumber
\end{equation}
is said to be completely integrable or Liouville integrable if
\begin{enumerate}
%\begin{itemize}
	\item there is a Poisson matrix $\left[J\right]_{ij}=\left\{x_i,x_j\right\}$, of rank $2r$, which is invariant under $Y$, namely the matrix $[\tilde{J}]_{ij}=\left\{u_i,u_j\right\}$ has the same functional form with $J$,
	\item map $Y$ has $r-$functionally independent invariants, which are in involution with respect to the corresponding Poisson bracket, i.e. $\left\{I_i,I_j\right\}=0$, $i,j=1,\ldots,r$, $i\neq j$,
	\item there are $k=2N-2r$ in the number Casimir functions, $C_i$, $i=1,\ldots,k$, which are invariant under $Y$, namely $C_i\circ Y=C_i$.
%\end{itemize}
\end{enumerate}
\end{CompleteIntegrability}

%%%%%%%%%%%%%%%%%%%%%%%%%%%%%%%%%%%%%%%%%%%%%%%%%%%%%%%%%%%%%%%%%%%%%%%%%%%%%%%%%%%%%%%%

\section{Organisation of the paper}
In the next section we briefly give some introduction to the notion of the \textit{reduction group} \cite{Mikhailov2}, \textit{automorphic Lie algebras} \cite{BuryPhD,Bury-Sasha,LombardoPhD,Lombardo} and \textit{Darboux trasformations} to make this text self-contained. We state three cases of reduction groups which are representative for all the \textit{finite reduction groups} with \textit{degenerate orbits}\cite{LombardoPhD,Lombardo}. 

In Section 4 we use Darboux transformations presented in \cite{SPS} to derive YB maps. In particular, we consider Darboux matrices for the NLS equation, the DNLS equation and for a deformation of the DNLS equation. For these Darboux matrices the refactorisation is not unique. Therefore, for the corresponding $6-$dimensional YB maps which are derived from the refactorisation problem, in principle, one needs to check the YB property separately. Yet, the entries of these Darboux matrices obey certain differential equations which possess first integrals. There is a natural restriction of the Darboux map on the affine variety corresponding to a level set of these first integrals. These restrictions make the refactorisation unique and this guarantees that the induced $4-$ dimensional YB maps satisfy the YB equation and they are reversible \cite{Veselov3}. We show that these YB maps have Poisson structure. However, the first integrals are not always very useful for the reduction because, in general, they are polynomial equations. In fact, in the case of the DNLS equation the corresponding 6-dimensional YB map cannot be reduced to a 4-dimensional one on the invariant leaves explicitly. However, after a change in the variables we obtain a reducible map. In the case of the deformation of the DNLS equation, we can only derive the $6-$dimensional YB map and its implicit restriction on $4-$dimensional invariant leaves. 

In Section 5 we consider the vector generalisations of the Adler-Yamilov YB map and the YB map corresponding to the DNLS equation.

%%%%%%%%%%%%%%%%%%%%%%%%%%%%%%%%%%%%%%%%%%%%%%%%%%%%%%%%%%%%%%%%%%%%%%%%%%%%%%%%%%%%%%%%%%%%%%%%%%%%%%%%%%%%%%%%%%%%%%%%%%%%%%%%%%%%%%%%

\section{Automorphic Lie algebras and Darboux Transformations and reduction groups with degenerate orbits}
The reduction group was first introduced in \cite{Mikhailov2}. It is a discrete group of automorphisms of a Lax operator and its elements are silmutaneous gauge transformations and fractional-linear transformations of the spectral parameter.

Automorphic Lie algebras were introduced in \cite{LombardoPhD,Lombardo} and studied in \cite{BuryPhD,Bury-Sasha,LombardoPhD,Lombardo}. These algebras constitute a subclass of infinite dimensional Lie algebras and their name is due to their construction which is very similar to the one for automorphic functions. 

Darboux transformations and their relations to the theory of \textit{integrable systems} have been extensively studied \cite{Matveev-Salle, Rog-Schief}. Such transformations can be derived from Lax pairs as, for instance, in \cite{Rog-Schief}, or in a more systematic algebraic manner in \cite{Cieslinski, SPS}.

We are interested in Darboux transformations corresponding to Lax operators of the following form
\begin{equation} \label{L-operator}
  \mathfrak{L}=\mathfrak{L}(\textbf{p}(x);\lambda)=D_{x}+U(\textbf{p}(x);\lambda),
\end{equation}
where $U$ belongs to an automorphic Lie algebra.

In the rest of the text we use ``$\mathfrak{L}$" for Lax operators and ``$L$" for Lax matrices of the refactorisation problem (\ref{eqLax}).

By Darboux transformations we understand maps 
\begin{equation}\label{DarbouxTransform}
	\mathfrak{L}(\textbf{p}(x);\lambda) \rightarrow \mathfrak{\widetilde{L}}:=\mathfrak{L}(\widetilde{\textbf{p}}(x);\lambda)=M\mathfrak{L}M^{-1},
\end{equation}
$M$ is a matrix called the \textit{Darboux matrix}. They map fundamental solutions, $\Psi$, of the equation $\mathfrak{L}\Psi=0$ to other fundamental solutions, $\widetilde{\Psi}=M\Psi$, of the equation $\mathfrak{\widetilde{L}}\widetilde{\Psi}=0$.

The structure of Lax operators has a natural Lie algebraic interpretation in terms of Kac-Moody algebras and automorphic Lie algebras \cite{BuryPhD,Bury-Sasha,LombardoPhD,Lombardo}. While a Kac-Moody algebra is associated with an automorphism of finite order, automorphic Lie algebras correspond to a finite group of automorphisms, namely the \textit{reduction group} \cite{Mikhailov2}.

In \cite{BuryPhD, Bury-Sasha} it was proved that in the case of $2\times 2$ matrices, which we study in this paper, the essentially different reduction groups are the trivial group (with no reduction), the cyclic group $\field{Z}_2$ (leading to the Kac-Moody algebra $A_1^1$) and the Klein group $\field{Z}_2\times \field{Z}_2$ \cite{Lombardo,MSY}.

We shall present 4 and $6-$dimensional YB maps for all the following cases. The trivial group associated with the NLS equation \cite{ZS}
\begin{equation}
p_t=p_{xx}+4p^2q,\qquad q_t=-q_{xx}-4pq^2.
\end{equation}
The $\field{Z}_2$ group associated to the DNLS equation equation \cite{Kaup-Newell}
\begin{equation}
p_t=p_{xx}+4(p^2q)_x,\qquad q_t=-q_{xx}+4(pq^2)_x.
\end{equation}
and the $\field{Z}_2\times \field{Z}_2$ group associated to the deformation of the DNLS equation \cite{MSY}
\begin{equation}
p_t=p_{xx}+8(p^2q)_x-4q_x,\qquad q_t=-q_{xx}+8(pq^2)_x-4p_x. 
\end{equation}

The above mentioned groups are representative of all the \textit{finite reduction groups} with \textit{degenerate orbits}, namely orbits corresponding to the fixed points of the fractional linear transformations of the spectral parameter.

%%%%%%%%%%%%%%%%%%%%%%%%%%%%%%%%%%%%%%%%%%%%%%%%%%%%%%%%%%%%%%%%%%%%%%%%%%%%%%%%
%%%%%%%%%%%%%%%%%%%%%%%%%%%%%%%%%%%%%%%%%%%%%%%%%%%%%%%%%%%%%%%%%%%%%%%%%%%%%%%%

\section{Derivation of YB maps}
In \cite{SPS} we used Darboux transformations to construct integrable systems of discrete equations, which have the multidimensional consistency property \cite{ABS-2004, ABS-2005, Bobenko-Suris,Frank3,Frank5,Frank4}. The compatibility condition of Darboux transformations around the square is exactly the same with the Lax equation (\ref{eqLax}). Therefore, in this paper, we use Darboux transformations to construct YB maps.

We start with the well known example of the Darboux transformation for the nonlinear Schr\"odinger equation and construct its associated YB map.

\subsection{The Nonlinear Schr\"odinger equation}
In the case of NLS equation, the Lax operator is given by
\begin{equation}
\mathfrak{L}(p,q;\lambda)=D_x+\lambda U_{1}+U_{0},\quad \text{where} \quad U_1=\sigma_3=\text{diag}(1,-1),\quad U_0=\left(\begin{matrix}
        0 & 2p \\
        2q & 0
    \end{matrix}\right).
\end{equation}

The Darboux Transformation, $M$, of $\mathfrak{L}$ is given by \cite{SPS, Rog-Schief}
\begin{equation}\label{NLSDarboux}
  M=\lambda \left(
     \begin{matrix}
         1 & 0\\
         0 & 0
     \end{matrix}\right)+\left(
     \begin{matrix}
         f & p\\
         \widetilde{q} & 1
     \end{matrix}\right).
\end{equation}
The entries of (\ref{NLSDarboux}), according to definition (\ref{DarbouxTransform}), must satisfy the following system of equations
\begin{equation}
\partial_x f=2(pq-\widetilde{p}\widetilde{q}), \qquad \partial_x p=2(pf-\widetilde{p}), \qquad \partial_x \widetilde{q}=2(q-\widetilde{q}f),\nonumber
\end{equation}
which admits the following first integral
\begin{equation}\label{integralNLS}
\partial_x(f-p\widetilde{q})=0.
\end{equation}
This integral implies that $\partial_x \det M=0$. 

In correspondence with (\ref{NLSDarboux}), we define the matrix
\begin{equation} \label{3d-Darboux-NLS}
  M(\textbf{x},X;\lambda)=\lambda \left(
     \begin{matrix}
         1 & 0\\
         0 & 0
     \end{matrix}\right)+\left(
     \begin{matrix}
         X & x_1\\
         x_2 & 1
     \end{matrix}\right)
\end{equation}
and substitute it into the Lax equation (\ref{eqLax}) 
\begin{equation}\label{laxM}
M(\textbf{u},U;\lambda)M(\textbf{v},V;\lambda)=M(\textbf{y},Y;\lambda)M(\textbf{x},X;\lambda),
\end{equation}
to derive the following system of equations
\begin{eqnarray}
&v_1 = x_1,\ u_2 = y_2,\ U +V = X + Y ,\ u_2 v_1 = x_1 y_2,& \nonumber \\
&u_1 +U v_1 = y_1+x_1 Y,\ u_1 v_2+U V = x_2 y_1+X Y,\ v_2+u_2 V = x_2 + X y_2.& \nonumber
\end{eqnarray}

The corresponding algebraic variety is a union of two six-dimensional components. The first one is obvious from the refactorisation problem (\ref{laxM}), and it corresponds to the permutation map
\begin{equation}
 \textbf{x}\mapsto \textbf{u}=\textbf{y}, \quad \textbf{y}\mapsto \textbf{v}=\textbf{x}, \quad X\mapsto U=Y, \quad Y\mapsto V=X, \nonumber
\end{equation}
which is a trivial YB map. The second one can be represented as a rational 6-dimensional non-involutive map of $K^3\times K^3 \rightarrow K^3\times K^3$

\begin{equation}\label{NLS-3d} 
 \begin{array}{ll}
x_1\mapsto u_1=\frac{y_1+x_1^2x_2-x_1X+x_1Y}{1+x_1y_2},& y_1\mapsto v_1=x_1, \\ 
x_2\mapsto u_2=y_2,&
y_2\mapsto v_2=\frac{x_2+y_1y_2^2+y_2X-y_2Y}{1+x_1y_2},  \\
X\mapsto U=\frac{y_1y_2-x_1x_2+X+x_1y_2Y}{1+x_1y_2},& Y\mapsto
V=\frac{x_1x_2-y_1y_2+x_1y_2X+Y}{1+x_1y_2},
 \end{array}
\end{equation}
which, one can easily check that, statisfies the YB equation.

It follows from (\ref{laxM}) that the trace of $M(\textbf{y},Y;\lambda)M(\textbf{x},X;\lambda)$, is a polynomial in $\lambda$ whose coefficients are
\begin{equation}
\mbox{Tr}(M(\textbf{y};b,\lambda)M(\textbf{x};a,\lambda))=\lambda^2+\lambda I_1(\textbf{x},\textbf{y},X,Y)+I_2(\textbf{x},\textbf{y},X,Y), \nonumber
\end{equation}
where
\begin{equation}\label{NLS3dInv}
I_1(\textbf{x},\textbf{y},X,Y)=X+Y \qquad \text{and} \qquad I_2(\textbf{x},\textbf{y},X,Y)=x_2y_1+x_1y_2+XY,
\end{equation}
and those are invariants for the YB map (\ref{NLS-3d}).

In the following section we show that the YB map (\ref{NLS-3d}) can be reduced to a $4-$ dimensional YB map which has Poisson structure.

%%%%%%%%%%%%%%%%%%%%%%%%%%%%%%%%%%%%%%%%%%%%%%%%%%%%%%%%%%%%%%%%%%%%%%%%%%%%%%%%%%%%

\subsubsection{Restriction on invariant leaves: The Adler-Yamilov map}
We now take into account the first integral (\ref{integralNLS}) of the Darboux matrix (\ref{NLSDarboux}). This first integral requires that the matrix entries $f$, $q$ and $\widetilde{q}$ are related as 
\begin{equation}
f-p\widetilde{q}=\alpha=constant. \nonumber
\end{equation}
Therefore, the entries of matrix (\ref{3d-Darboux-NLS}) must obey the following equation
\begin{equation}\label{X-relation}
X-x_1x_2=a,
\end{equation}
respectively.

The matrix (\ref{3d-Darboux-NLS}) now takes the form
\begin{equation} \label{laxNLS}
M(\textbf{x};a,\lambda)=\lambda \left(
\begin{matrix}
 1 & 0\\
 0 & 0
\end{matrix}\right)+\left(
\begin{matrix}
 a+x_1x_2 & x_1\\
 x_2 & 1
\end{matrix}\right).
\end{equation}
In this case, the Lax equation (\ref{eqLax}),
\begin{equation} \label{lax_eq_NLS}
  M(\textbf{u};a,\lambda)M(\textbf{v};b,\lambda)=M(\textbf{y};b,\lambda)M(\textbf{x};a,\lambda),
\end{equation}
has a unique solution $\textbf{u}=\textbf{u}(\textbf{x},\textbf{y})$, $\textbf{v}=\textbf{v}(\textbf{x},\textbf{y})$ which defines a map $\textbf{x}\rightarrow \textbf{u}(\textbf{x},\textbf{y})$, $\textbf{y}\rightarrow \textbf{v}(\textbf{x},\textbf{y})$, given by
\begin{eqnarray} \label{YB_NLS}
(\textbf{x},\textbf{y})\overset{Y_{a,b}}{\longrightarrow }\left(y_1-\frac{a -b}{1+x_1y_2}x_1,y_2,x_1,x_2+\frac{a -b}{1+x_1y_2}y_2\right).
\end{eqnarray}
Therefore, the above map is a reversible parametric YB map with strong Lax matrix ($\ref{laxNLS}$). Moreover, it is not involutive.

The map (\ref{YB_NLS}) first appeared in the work of Adler Yamilov \cite{Adler-Yamilov}. Its interpretation as a YB map was given in \cite{Kouloukas, PT}.

From the trace of $M(\textbf{y};b,\lambda)M(\textbf{x};a,\lambda)$ we obtain the following invariants for the map (\ref{YB_NLS})
\begin{eqnarray}
&&I_1(\textbf{x},\textbf{y})=x_1 x_2+y_1 y_2+a+b, \\
&&I_2(\textbf{x},\textbf{y})=(a+x_1 x_2)(b+y_1 y_2)+x_1 y_2+x_2 y_1+1.
\end{eqnarray}
The constant terms in $I_1,I_2$ can be omitted. It is easy to check that $I_1,I_2$ are in involution with respect to invariant Poisson brackets defined as
\begin{equation}
\left\{x_1,x_2\right\}=\left\{y_1,y_2\right\}=1,\qquad \text{and all the rest}
\qquad \left\{x_i,y_j\right\}=0,
\end{equation}
and the corresponding Poisson matrix is invariant under the YB map (\ref{YB_NLS}). Therefore the map (\ref{YB_NLS}) is completely integrable.

The Adler-Yamilov map is a restriction of the YB map (\ref{NLS-3d}) on the invariant leaves 
\begin{equation}
A_a=\{(x_1,x_2,X)\in \field{R}^3; X=a+x_1x_2\}, \quad B_b=\{(y_1,y_2,Y)\in \field{R}^3; Y=b+y_1y_2\}.
\end{equation}

%%%%%%%%%%%%%%%%%%%%%%%%%%%%%%%%%%%%%%%%%%%%%%%%%%%%%%%%%%%%%%%%%%%%%%%%%%%%%%%%
%%%%%%%%%%%%%%%%%%%%%%%%%%%%%%%%%%%%%%%%%%%%%%%%%%%%%%%%%%%%%%%%%%%%%%%%%%%%%%%%

\subsection{Derivative NLS equation: $\field{Z}_2$ reduction}
The Lax operator for the DNLS equation \cite{CLH, Kaup-Newell} is given by
\begin{equation}
   \mathfrak{L}(p,q;\lambda)=D_x+\lambda^{2} U_2+\lambda U_1,\qquad \text{where} \qquad U_2=\sigma_3,\qquad U_1=\left(\begin{matrix} 0 & 2p \\ 2q & 0\end{matrix}\right),
\end{equation}
and $\sigma_3$ is a Pauli matrix. Operator $\mathfrak{L}$ is invariant with respect to the following involution
\begin{equation}\label{sym_cond}
  \mathfrak{L}(\lambda)=\sigma_{3}\mathfrak{L}(-\lambda) \sigma_{3},
\end{equation}
where $\mathfrak{L}(\lambda)\equiv \mathfrak{L}(p,q;\lambda)$. Involution ($\ref{sym_cond}$) generates the so-called reduction group \cite{Mikhailov2,Lombardo} and it is isomorphic to $\field{Z}_2$. 

The Darboux matrix in this case is given by \cite{SPS}
\begin{equation} \label{affDarboux}
  M:=\lambda^{2}\left(\begin{matrix}
    f & 0\\
    0 & 0\end{matrix}\right)+\lambda\left(\begin{matrix}
    0 & f p\\
    f\widetilde{q} & 0\end{matrix}\right)+\left(\begin{matrix}
    c & 0\\
    0 & 1\end{matrix}\right),
\end{equation} 
whose entries $p$, $\widetilde{q}$ and $f$ obey the following system of equations
%\numparts
\begin{eqnarray}\label{baecklundZ2-a}
\partial_x p&=&2p(\widetilde{p}\widetilde{q}-pq)-\frac{2}{f}(\widetilde{p}-cp), \\
\label{baecklundZ2-b}
\partial_x \widetilde{q}&=&2\widetilde{q}(\widetilde{p}\widetilde{q}-pq)-\frac{2}{f}(c\widetilde{q}-q), \\
\label{baecklundZ2-c}
\partial_x f&=&2f(pq-\widetilde{p}\widetilde{q}). 
\end{eqnarray}
%\endnumparts
System (\ref{baecklundZ2-a})-(\ref{baecklundZ2-c}) has a first integral which obliges the determinant of matrix (\ref{affDarboux}) to be $x-$independent, and it is given by
\begin{equation}\label{alpha-eq}
\partial_x (f^{2}p\widetilde{q}-f)=0.
\end{equation}

Using the entries of (\ref{affDarboux}) as variables, namely $(p,\widetilde{q},f;c)\rightarrow (x_1,x_2,X;1)$, we define the matrix
\begin{equation}\label{affDarboux-3D}
  M(\textbf{x},X;\lambda)=\lambda^{2}\left(\begin{matrix}
    X & 0\\
    0 & 0\end{matrix}\right)+\lambda\left(\begin{matrix}
    0 & x_1X\\
    x_2X & 0\end{matrix}\right)+\left(\begin{matrix}
    1 & 0\\
    0 & 1\end{matrix}\right).
\end{equation} 

The Lax equation implies the following equations
\begin{equation} 
 \begin{array}[pos]{ccc}
u_1U+v_1V=x_1X+y_1Y, \quad u_2U+v_2V=x_2X+y_2Y,\\ 
UV=XY, \quad v_1UV=x_1XY, \quad u_2UV=y_2XY, \quad u_2v_1UV=x_1y_2XY,  \\
U+V+u_1v_2UV=X+Y+x_2y_1XY.   
\end{array}
\end{equation}

As in the case of nonlinear Schr\"odinger equation, the algebraic variety consists of two components. The first 6-dimensional component corresponds to the permutation map
\begin{equation} \label{perMap}
 \textbf{x}\mapsto \textbf{u}=\textbf{y}, \quad \textbf{y}\mapsto \textbf{v}=\textbf{x}, \quad X\mapsto U=Y, \quad Y\mapsto V=X, 
\end{equation}
and the second corresponds to the following 6-dimensional YB map
\begin{equation}\label{f-g-h}
\begin{array}[pos]{lcl}
x_1 \mapsto u_1=f_1(\textbf{x},\textbf{y},X,Y),\qquad y_1 \mapsto v_1=f_2(\pi\textbf{y},\pi\textbf{x},Y,X), \\ 
x_2 \mapsto u_2=f_2(\textbf{x},\textbf{y},X,Y),\qquad y_2 \mapsto v_2=f_1(\pi\textbf{y},\pi\textbf{x},Y,X), \\ 
\,X \mapsto U=f_3(\textbf{x},\textbf{y},X,Y),\qquad \, Y \mapsto V=f_3(\pi\textbf{y},\pi\textbf{x},Y,X),
\end{array}
\end{equation}
where $\pi$ is the \textit{permutation function}, $\pi(x_1,x_2)=(x_2,x_1)$, $\pi^2=1$ and $f_1,f_2$ and $f_3$ are given by
%\numparts
\begin{eqnarray} \label{f1-Aff}
\hspace{-0.5cm} f_1(\textbf{x},\textbf{y},X,Y)&=&\frac{-1}{f_3(\textbf{x},\textbf{y})}\frac{x_1X+(y_1-x_1)Y-x_1x_2y_1XY-x_1^2x_2X^2}{x_1 x_2 X+x_1 y_2 Y-1}, \\
\label{f2-Aff}
\hspace{-0.5cm} f_2(\textbf{x},\textbf{y},X,Y)&=&y_2, \\ 
\label{f3-Aff}
\hspace{-0.5cm} f_3(\textbf{x},\textbf{y},X,Y)&=&\frac{x_1x_2X+x_1y_2Y-1}{x_1y_2X+y_1 y_2 Y-1}X.
\end{eqnarray}
%\endnumparts

One can verify that the above map is a non-involutive YB map. The invariants of this map are given by 
\begin{equation}
I_1(\textbf{x}\cdot\pi\textbf{y},X,Y)=XY\qquad \text{and}\qquad I_2(\textbf{x},\textbf{y},X,Y)=(\textbf{x}\cdot\pi\textbf{y})XY+X+Y.
\end{equation}

However, this map cannot be easily reduced on invariant leaves as in the case of NLS equation. This is due to the fact that the first integral of system (\ref{baecklundZ2-a})-(\ref{baecklundZ2-c}) is
\begin{equation}
f-f^2p\widetilde{q}=k,
\end{equation}
and therefore the relation between the entries $x_1, x_2$ and $X$ of (\ref{affDarboux-3D}) reads
\begin{equation}
X-X^2x_1x_2=k.
\end{equation}
The solution for $X$ of the above equation can only be expressed in terms of square roots of $x_1$ and $x_2$. Thus, in this case is difficult to solve the Lax equation, and if we do so the resulting map will not be presentable because of its length.

One option is to present the YB map implicitly. Specifically, the corresponding $4-$dimensional map is given by
\begin{equation}
\begin{array}[pos]{lcl}
x_1 \mapsto u_1=f_1(\textbf{x},\textbf{y},X,Y),\qquad y_1 \mapsto v_1=f_2(\pi\textbf{y},\pi\textbf{x},Y,X), \\ 
x_2 \mapsto u_2=f_2(\textbf{x},\textbf{y},X,Y),\qquad y_2 \mapsto v_2=f_1(\pi\textbf{y},\pi\textbf{x},Y,X), 
\end{array}
\end{equation}
where $f_i$, $i=1,2$, are given  by (\ref{f1-Aff})-(\ref{f2-Aff}) where $X$ and $Y$ are given by
\begin{equation}
X-X^2x_1x_2=a, \qquad Y-Y^2y_1y_2=b.
\end{equation}

Nevertheless, we can choose another parametrisation of matrix (\ref{affDarboux}), in order to make the obtained YB map explicitly reducible on the invariant leaves.

%%%%%%%%%%%%%%%%%%%%%%%%%%%%%%%%%%%%%%%%%%%%%%%%%%%%%%%%%%%%%%%%%%%%%%%%%%%%%%%%%%%%%%%%%%%%%%%%%%

\subsubsection{$\field{Z}_2$ reduction: A reducible 6-dimensional YB map} 
Now, lets go back to the Darboux matrix (\ref{affDarboux}) and  replace $(f p,f \widetilde{q},f;c)\rightarrow (x_1,x_2,X;1)$, namely  
\begin{equation}\label{Affine-3d1}
  M(\textbf{x},X;\lambda)=\lambda^{2}\left(\begin{matrix}
    X & 0\\
    0 & 0\end{matrix}\right)+\lambda\left(\begin{matrix}
    0 & x_1\\
    x_2 & 0\end{matrix}\right)+\left(\begin{matrix}
    1 & 0\\
    0 & 1\end{matrix}\right),
\end{equation} 
 
From the Lax equation we obtain the following equations
\begin{eqnarray}
u_2v_1=x_1y_2, \quad u_2V=Xy_2, \quad Uv_1=x_1Y, \quad UV=XY \nonumber \\
u_1+v_1=x_1+y_1, \quad U+u_1v_2+V=X+x_2y_1+Y, \quad u_2+v_2=x_2+y_2. \nonumber
\end{eqnarray}

Now, the first 6-dimensional component of the algebraic variety corresponds to the trivial map (\ref{perMap}) and the second component corresponds to a map of the form (\ref{f-g-h}), with $f_1,f_2$ and $f_3$ now given by
%\numparts
\begin{eqnarray} \label{YB-Aff-3d1}
f_1(\textbf{x},\textbf{y},X,Y)&=&\frac{(x_1+y_1)X-x_1Y-x_1x_2(x_1+y_1)}{X-x_1(x_2+y_2)}, \\
\label{YB-Aff-3d2}
f_2(\textbf{x},\textbf{y},X,Y)&=&\frac{X-x_1(x_2+y_2)}{Y-y_2(x_1+y_1)}y_2,  \\ 
\label{YB-Aff-3d3}
f_3(\textbf{x},\textbf{y},X,Y)&=&\frac{X-x_1(x_2+y_2)}{Y-y_2(x_1+y_1)}Y. 
\end{eqnarray}
%\endnumparts

This map has the following invariants
\begin{eqnarray}\label{invar-aff-3d}
I_1(\textbf{x},\textbf{y},X,Y)=XY, \qquad \quad \,\,I_2(\textbf{x},\textbf{y},X,Y)=\textbf{x}\cdot\pi\textbf{y}+X+Y,\\
\label{invar-aff-3d-2}
I_3(\textbf{x},\textbf{y},X,Y)=x_1+y_1, \qquad I_4(\textbf{x},\textbf{y},X,Y)=x_2+y_2.
\end{eqnarray}

%%%%%%%%%%%%%%%%%%%%%%%%%%%%%%%%%%%%%%%%%%%%%%%%%%%%%%%%%%%%%%%%%%%%%%%%%%%%%%%%%%%%%%%%%%%%%%%%%%%%%%

\subsubsection{Restriction on invariant leaves}
Now, taking into account the first integral 
\begin{equation}
f-(fp)(f\widetilde{q})=k=constant,
\end{equation}
the entries of matrix (\ref{Affine-3d1}) must satisfy the following equation
\begin{equation}\label{const.Det.}
X-x_1x_2=k,
\end{equation}
and thus matrix (\ref{Affine-3d1}) takes the following form
\begin{equation} \label{Darboux_Affine}
  M(\textbf{x};k;\lambda)=\lambda^{2}\left(\begin{matrix}
    k+x_1x_2 & 0\\
    0 & 0\end{matrix}\right)+\lambda\left(\begin{matrix}
    0 & x_1\\
    x_2 & 0\end{matrix}\right)+\left(\begin{matrix}
    1 & 0\\
    0 & 1\end{matrix}\right).
\end{equation} 

The Lax-equation for matrix (\ref{Darboux_Affine}) is equivalent to the following
\begin{equation} \label{YB-affine}
\left(\textbf{x},\textbf{y}\right)\overset{Y_{a,b}}{\longrightarrow }\left(y_1+\frac{a-b }{a-x_1y_2}x_1, \frac{a-x_1 y_2}{b-x_1 y_2}y_2, \frac{b-x_1 y_2}{a-x_1 y_2}x_1,x_2+\frac{b-a}{b-x_1 y_2}y_2\right),
\end{equation}
and therefore the above map is a reversible parametric YB map. Moreover, it is not involutive.

The invariants of map (\ref{YB-affine}) are given by
\begin{equation}
I_1(\textbf{x},\textbf{y})=(a+x_1x_2)(b+y_1y_2), \qquad I_2(\textbf{x},\textbf{y})=(x_1+y_1)(x_2+y_2)+a+b.
\end{equation}
The constant terms in $I_1$ and $I_2$ can be omitted. These are the invariants we retrieve from the trace of $M(\textbf{y};b,\lambda)M(\textbf{x};a,\lambda)$. However, the quantities $x_1+y_1$ and $x_2+y_2$ in $I_2$ are invariants themselves. The Poisson bracket in this case is given by
\begin{equation}
\left\{x_1,x_2\right\}=\left\{y_1,y_2\right\}=\left\{x_2,y_1\right\}=\left\{y_2,x_1\right\}=1,\quad \text{and all the rest} \quad \left\{x_i,y_j\right\}=0.
\end{equation}
The rank of the Poisson matrix is 2, $I_1$ is one invariant and $I_2=C_1C_2+a+b$, where $C_1=x_1+y_1$ and $C_2=x_2+y_2$ are Casimir functions. The latter are preserved by (\ref{YB-affine}), namely $C_i\circ Y_{a,b}=C_i$, $i=1,2$. Therefore, map (\ref{YB-affine}) is completely integrable.

Map (\ref{YB-affine}) is a restriction of the YB map (\ref{YB-Aff-3d1})-(\ref{YB-Aff-3d3}) on the invariant leaves 
\begin{equation}
A_a=\{(x_1,x_2,X)\in \field{R}^3; X=a+x_1x_2\}, \quad B_b=\{(y_1,y_2,Y)\in \field{R}^3; Y=b+y_1y_2\}.
\end{equation}

Moreover, the map (\ref{YB-affine}) can be expressed as a map of two variables on the symplectic leaf
\begin{equation}
x_1+y_1=c_1, \qquad x_2+y_2=c_2.
\end{equation}

%%%%%%%%%%%%%%%%%%%%%%%%%%%%%%%%%%%%%%%%%%%%%%%%%%%%%%%%%%%%%%%%%%%%%%%%%%%%%%%%
%%%%%%%%%%%%%%%%%%%%%%%%%%%%%%%%%%%%%%%%%%%%%%%%%%%%%%%%%%%%%%%%%%%%%%%%%%%%%%%%
\subsection{A deformation of the DNLS equation: Dihedral Group}
In the case of dihedral reduction group, the Lax operator is given by
\begin{equation}\label{laxDihedral}
\begin{array}[pos]{ccl}
 \mathfrak{L}(p,q;\lambda)=D_x+\lambda^{2}U_2+\lambda U_1+\lambda^{-1} U_{-1}-\lambda^{-2}U_{-2},\qquad \text{where} \\ 
 U_{2}\equiv U_{-2}=\sigma_3, \qquad 
 U_1=\left(\begin{matrix} 0 & 2p\\ 2q & 0\end{matrix}\right) \qquad U_{-1}=\sigma_1 U_1 \sigma_1,
\end{array}
\end{equation}
and $\sigma_1$, $\sigma_3$ are Pauli matrices. Here, the reduction group consists of the following set of transformations acting on the Lax operator (\ref{laxDihedral}) ,
\begin{equation}
 \mathfrak{L}(\lambda)=\sigma_3\mathfrak{L}(-\lambda)\sigma_3 \qquad \text{and} \qquad \mathfrak{L}(\lambda)
 =\sigma_1\mathfrak{L}(\lambda^{-1})\sigma_1,
\label{reduct_group}
\end{equation}
and it is isomorphic to $\field{Z}_2\times\field{Z}_2\cong D_2$, \cite{Lombardo}.

In this case, the Darboux matrix is given by \cite{SPS}
\begin{equation} \label{Darboux-D2}
  M=f\left(\left(\begin{matrix} 
    \lambda^2 & 0\\
    0 & \lambda^{-2}\end{matrix}\right)+\lambda\left(\begin{matrix}
    0 & p\\
    \widetilde{q} & 0\end{matrix}\right)+
    g\left(\begin{matrix} 
    1 & 0\\
    0 & 1\end{matrix}\right)
    +\frac{1}{\lambda}\left(\begin{matrix}
    0 & \widetilde{q}\\
    p & 0\end{matrix}\right)\right),
\end{equation}
where its entries obey the following equations
%\numparts
\begin{eqnarray}
\partial_x p=2((\widetilde{p}\widetilde{q}-pq)p+(p-\widetilde{p})g+q-\widetilde{q}), \\
\partial_x \widetilde{q}=2((\widetilde{p}\widetilde{q}-pq)\widetilde{q}+p-\widetilde{p}+(q-\widetilde{q})g), \\
\partial_x g=2((\widetilde{p}\widetilde{q}-pq)g+(p-\widetilde{p})p+(q-\widetilde{q})\widetilde{q}), \\
\partial_x f=-2(\widetilde{p}\widetilde{q}-pq))f.
\end{eqnarray}
%\endnumparts
It can be shown that the above system of differential equations admits two first integrals, $\partial_x\Phi_i=0$, $i=1,2$, where
\begin{equation}\label{eq.f-g}
  \Phi_1:=f^2(g-p\widetilde{q}) \qquad \text{and}\qquad
  \Phi_2:=f^2(g^2+1-p^2-\widetilde{q}^2).
\end{equation}

It follows from (\ref{eq.f-g}), that the quantities $f$ and $fg$ can be expressed in terms of $p$ and $\widetilde{q}$, as solutions of quadratic equations. Then, the Darboux matrix depends only on two variables and we construct a $4-$dimensional parametric YB map. Although, we have omitted this map because of its length. 

In the next section we construct a $6-$dimensional map from (\ref{Darboux-D2}).

\subsubsection{Dihedral group: A 6-dimensional YB map}
We now consider the matrix $N:=fM$, where $M$ is given by (\ref{Darboux-D2}), and we change $(p,\widetilde{q},f^2)\rightarrow (x_1,x_2,X)$. Then,
\begin{equation} \label{3dLaxD2}
  N(\textbf{x},X;c_1,\lambda)=\left(\begin{matrix}
    \lambda^{2}X+x_1x_2X+c_1 & \lambda x_1X+\lambda^{-1}x_2X\\
    \lambda x_2X+\lambda^{-1}x_1X & \lambda^{-2}X+x_1x_2X+c_1\end{matrix}\right),
\end{equation}
where we have substituted the product $f^2g$ by
\begin{equation}\label{f-g-X}
f^2g=c_1+x_1x_2X,
\end{equation}
using the first integral, $\Phi_1$, in (\ref{eq.f-g}).

The Lax equation for the Darboux matrix (\ref{3dLaxD2}) reads
\begin{equation}
N(\textbf{u},U;a,\lambda)N(\textbf{v},V;b,\lambda)=N(\textbf{y},Y;b,\lambda)N(\textbf{x},X;a,\lambda),
\end{equation}
from where we obtain an algebraic system of equations, omitted because of its length.

The first 6-dimensional component of the corresponding algebraic variety corresponds to the trivial YB map 
\begin{equation} 
 \textbf{x}\mapsto \textbf{u}=\textbf{y}, \quad \textbf{y}\mapsto \textbf{v}=\textbf{x}, \quad X\mapsto U=\frac{a}{b}Y, \quad Y\mapsto V=\frac{b}{a}X, \nonumber
\end{equation}
and the second component corresponds to the following map
\begin{eqnarray}\label{f-g-h-D2}
&&x_1 \mapsto u_1=\frac{f(\textbf{x},\textbf{y},X,Y;a,b)}{g(\textbf{x},\textbf{y},X,Y;a,b)},\qquad y_1 \mapsto v_1=x_1 \nonumber \\ 
&&x_2 \mapsto u_2=y_2,\qquad \qquad \qquad \qquad ~ y_2 \mapsto v_2=\frac{f(\pi\textbf{y},\pi\textbf{x},Y,X;b,a)}{g(\pi\textbf{y},\pi\textbf{x},Y,X;b,a)} \\ 
&&~X \mapsto U=\frac{g(\textbf{x},\textbf{y},X,Y;a,b)}{h(\textbf{x},\textbf{y},X,Y;a,b)},\qquad Y \mapsto V=\frac{g(\pi\textbf{y},\pi\textbf{x},Y,X;b,a)}{h(\pi\textbf{y},\pi\textbf{x},Y,X;b,a)}\nonumber,
\end{eqnarray}
where $f,g$ and $h$ are given by
\begin{equation}
\begin{array}[pos]{lcl}
&&f(\textbf{x},\textbf{y},X,Y;a,b)=a^2b^2x_1X+a^2b[x_2-y_2+2x_1x_2y_1+x_1^2(y_2-3x_2)]XY+\\
&&\qquad a^2(y_2^2-1)[y_1(1+x_1^2)-x_1(1+y_1^2)]XY^2-ab^2(x_1^2-1)(y_2-x_2)X^2-\\
&&\qquad ab(x_1^2-1)[x_2^2(3x_1-y_1)-x_1-y_1+2y_2(y_1y_2-x_1x_2)]X^2Y-\\
&&\qquad a(x_1^2-1)(y_2^2-1)[y_2(y_1^2-1)+x_2(y_1^2-2x_1y_1+1)]X^2Y^2+\\
&&\qquad y_1(x_1^2-1)^2(x_2^2-1)(y_2^2-1)X^3Y^2+b(x_1^2-1)^2(x_2^2-1)(y_2-x_2)X^3Y+\\
&&\qquad a^3b(y_1-x_1)Y,\\ \\
&&g(\textbf{x},\textbf{y},X,Y;a,b)=a^2b^2X+2a^2by_2(y_1-x_1)XY+a^2(y_2^2-1)(x_1-y_1)^2XY^2+\\
&&\qquad 2ab(x_1^2-1)(1-x_2y_2)X^2Y+2ax_2(x_1^2-1)(y_2^2-1)(x_1-y_1)X^2Y^2+\\
&&\qquad (x_1^2-1)^2(x_2^2-1)(y_2^2-1)X^3Y^2,\\ \\
&&h(\textbf{x},\textbf{y},X,Y;a,b)=a^2b^2-2ab^2x_1(y_2-x_2)X-2ab(x_1y_1-1)(y_2^2-1)XY\\
&&\qquad b^2(x_1^2-1)(x_2-y_2)^2X^2-2by_1(x_2-y_2)(x_1^2-1)(y_2^2-1)X^2Y+\\
&&\qquad (x_1^2-1)(y_1^2-1)(y_2^2-1)^2X^2Y^2.
\end{array}
\end{equation}

It can be verified that this is a parametric YB map. From $\text{Tr}(N(\textbf{x},X;\lambda)N(\textbf{y},Y;\lambda))$ we extract the following invariants for the above map
%\numparts
\begin{eqnarray}
\hspace{-1.2cm} I_1(\textbf{x},\textbf{y},X,Y;a,b)&=&XY,  \\
\hspace{-1.2cm} I_2(\textbf{x},\textbf{y},X,Y;a,b)&=&bX+aY+(x_1+y_1)(x_2+y_2)XY,  \\
\hspace{-1.2cm} I_3(\textbf{x},\textbf{y},X,Y;a,b)&=&2bx_1x_2X+2ay_1y_2Y+2(\textbf{x}\cdot\textbf{y}+x_1x_2y_1y_2)XY+2ab. 
\end{eqnarray}
%\endnumparts

\subsubsection{Restriction to invariant leaves: An implicit map}
Using the first integrals $\Phi_1$ and $\Phi_2$ we can reduce the above $6-$dimensional map to a $4-$dimensional YB map implicitly. In particular, from equations $\Phi_1=c_1$ and $\Phi_2=c_2$ one can obtain
\begin{equation}
(1-x_1^2-x_2^2+x_1^2x_2^2)X^2+(2x_1x_2-c_2)X+1=0,
\end{equation}
where we have rescaled $c_1\rightarrow 1$.

Therefore the $4-$dimensional map is given by
\begin{equation}
(u_1,u_2,v_1,v_2)=\left(\frac{f(\textbf{x},\textbf{y},X,Y;a,b)}{g(\textbf{x},\textbf{y},X,Y;a,b)},y_2,x_1,\frac{f(\pi\textbf{y},\pi\textbf{x},X,Y;b,a)}{g(\pi\textbf{y},\pi\textbf{x},X,Y;b,a)}\right),
\end{equation}
where $f$, $g$ and $h$ are given by the above relations and $X$ and $Y$ are given by
%\numparts
\begin{eqnarray}
(1-x_1^2-x_2^2+x_1^2x_2^2)X^2+(2x_1x_2-a)X+1&=&0, \\
(1-y_1^2-y_2^2+y_1^2y_2^2)Y^2+(2y_1y_2-b)Y+1&=&0.
\end{eqnarray}
%\endnumparts

\subsection{Dihedral group: A linearised YB map}
We replace $(f \widetilde{q},f p)\rightarrow (x_1,x_2)$ in the Darboux matrix (\ref{Darboux-D2}) to become
\begin{equation}\label{Lax_D2}
  M(\textbf{x};k,\lambda)=\left(\begin{matrix} 
    \lambda^2 f & 0\\
    0 & \lambda^{-2} f\end{matrix}\right)+\lambda\left(\begin{matrix}
    0 & x_1\\
    x_2 & 0\end{matrix}\right)+
    fg\left(\begin{matrix}
    1 & 0\\
    0 & 1\end{matrix}\right)
    +\frac{1}{\lambda}\left(\begin{matrix}
    0 & x_2\\
    x_1 & 0\end{matrix}\right).
\end{equation}
Then equations $\Phi_1=c_1$ and $\Phi_2=c_2$ imply that the quantities $f$ and $fg$ are given by
%\numparts
\begin{eqnarray}
 f&=&\frac{1}{2} \sqrt{1+(x_1+x_2)^2}+\frac{1}{2} \sqrt{k^2+(x_1-x_2)^2},\\ fg&=&\frac{1}{2} \sqrt{1+(x_1+x_2)^2}-\frac{1}{2} \sqrt{k^2+(x_1-x_2)^2},
\end{eqnarray}
%\endnumparts
where $(c_1,c_2)\rightarrow (\frac{1-k^2}{4},\frac{1+k^2}{2})$.

The linear approximation to the YB map is given by
\begin{equation}
\left(\begin{matrix} x_1 \\ x_2 \\ y_1\\ y_2 \end{matrix}\right) \overset{U_0}{\longrightarrow}
\left(\begin{matrix} u_1 \\ u_2 \\ v_1\\ v_2 \end{matrix}\right)=
\left(\begin{array}{c c|c c}
\frac{(a-1)(a-b)}{(a+1)(a+b)} & \frac{a-b}{a+b} & \frac{2 a}{a+b} & \frac{(a+1)(b-a)}{(b+1)(a+b)} \\
0 & 0 & 0 & \frac{a+1}{b+1}\\
\hline
\frac{b+1}{a+1} & 0 & 0 & 0\\
\frac{(a-b)(b+1)}{(a+1)(a+b)} & \frac{2b}{a+b} & \frac{b-a}{a+b} & \frac{(b-1)(b-a)}{(b+1)(a+b)}\end{array}\right)
\left(\begin{matrix} x_1 \\ x_2 \\ y_1\\ y_2 \end{matrix}\right)
\label{linear_YB_D2}
\end{equation}
which is a linear parametric YB map and it is not involutive.

%%%%%%%%%%%%%%%%%%%%%%%%%%%%%%%%%%%%%%%%%%%%%%%%%%%%%%%%%%%%%%%%%%%%%%%%%%%%%%%%%%%%%%%%%%%%%%%%%%%%%%%%%%%%%%%%%%%%%%%%%%%%%%%%%%%%%%%%%%%%%

\section{$2N\times 2N-$dimensional YB maps}
In this section we consider the vector generalisations of the YB maps (\ref{YB_NLS}) and (\ref{YB-affine}).We replace the variables, $x_1$ and $x_2$, in the Lax matrices with $N-$vectors $\textbf{w}_1$ and $\textbf{w}_2^T$ to obtain $2N \times 2N$ YB maps. In what follows we use the following notation for a $n-$vector $\textbf{w}=(w_1,...,w_n)$
\begin{equation}
\textbf{w}=(\textbf{w}_1,\textbf{w}_2),\qquad \text{where} \qquad \textbf{w}_1=(w_1,...,w_N), \qquad \textbf{w}_2=(w_{N+1},...,w_{2N})
\end{equation}
and also
\begin{equation}
\langle u_i|:=\textbf{u}_i,\qquad |w_i\rangle:=\textbf{w}_i^T \qquad \text{and their dot product with}\qquad \langle u_i,w_i\rangle.
\end{equation}

\subsection{NLS equation}
Replacing the variables in (\ref{laxNLS}) with $N-$vectors, namely
\begin{equation}
M(\textbf{w};a,\lambda)=\left(     
\begin{matrix}
\lambda+a+\langle w_1,w_2\rangle  & \langle w_1| \\
|w_2\rangle      & I             
\end{matrix}\right),
\label{laxNLSnN}
\end{equation}
we obtain a unique solution of the Lax-Equation given by the following $2N\times 2N$ map
%\numparts
\begin{equation}
\begin{cases}
\langle u_1|=\langle y_1|+f(z;a,b)\langle x_1|,\\
\langle u_2|=\langle y_2|,
\end{cases}
\end{equation}
and
\begin{equation}
\begin{cases}
\langle v_1|=\langle x_1|,\\
\langle v_2|=\langle x_2+f(z;b,a)\langle y_2|,
\end{cases}
\end{equation}
where $f$ is given by
\begin{equation}
f(z;b,a)=\frac{b-a}{1+z},\qquad z:=\langle x_1,y_2\rangle.
\end{equation}
%\endnumparts
The above is a non-involutive parametric $2N\times 2N$ YB map with strong Lax matrix given by (\ref{laxNLSnN}). As a YB map it appears in \cite{PT}, but it is originally introduced by Adler \cite{AdlerNLSVec}. Moreover, one can construct the above $2N\times 2N$ map for the $N\times N$ Darboux matrix (\ref{laxNLSnN}) by taking the limit of the solution of the refactorisation problem in \cite{Kouloukas2}.

Two invariants of this map are given by
%\numparts
\begin{eqnarray}
\hspace{-1cm}I_1(\textbf{x},\textbf{y};a,b)=\langle x_1,x_2\rangle+\langle y_1,y_2\rangle,&\\ 
\hspace{-1cm}I_2(\textbf{x},\textbf{y};a,b)=b\langle x_1,x_2\rangle+a\langle y_1,y_2\rangle+\langle x_1,y_2\rangle+\langle x_2,y_1\rangle+\langle x_1,x_2\rangle \langle y_1,y_2\rangle.
\end{eqnarray}
%\endnumparts
These are the invariants which are obtained from the trace of $M(\textbf{y};b,\lambda)M(\textbf{x};a,\lambda)$ and they are not enough to prove Liouville integrability.
%%%%%%%%%%%%%%%%%%%%%%%%%%%%%%%%%%%%%%%%%%%%%%%%%%%%%%%%%%%%%%%%%%%%%%%%%%%%%%%%%%%%%%

\subsection{$\field{Z}_2$ reduction}
In the case of $Z_2$ we consider, instead of (\ref{Darboux_Affine}), the following matrix
\begin{equation}
M(\textbf{w};a,\lambda)=\left(     
\begin{matrix}
\lambda^2(a+\langle w_1,w_2\rangle)  & \lambda \langle w_1|\\
\lambda |w_2\rangle      & I           
\end{matrix}\right),
\label{laxAffinenN}
\end{equation}
we obtain a unique solution for the Lax-Equation given by the following $2N\times 2N$ map
%\numparts
\begin{equation}
\begin{cases}
\langle u_1|=\langle y_1|+f(z;a,b)\langle x_1|, \\
\langle u_2|=g(z;a,b)\langle y_2|,
\end{cases}
\end{equation}
and
\begin{equation}
\begin{cases}
\langle v_1|=g(z;b,a)\langle x_1|,\\
\langle v_2|=\langle x_2|+f(z;b,a)\langle y_2|,
\end{cases}
\end{equation}
where $f$ and $g$ are given by
\begin{equation}
f(z;a,b)=\frac{a-b}{a-z},\qquad g(z;a,b)=\frac{a-z}{b-z},\qquad z:=\langle x_1,y_2\rangle.
\end{equation}
%\endnumparts

The above map is a non-involutive parametric $2N\times 2N$ YB map with strong Lax matrix given by (\ref{laxAffinenN}).

The invariants of the above map, coming from the trace of $M(\textbf{y};b,\lambda)M(\textbf{x};a,\lambda)$, are given by
%\numparts
\begin{eqnarray}
I_1(\textbf{x},\textbf{y};a,b)&=&\langle x_1+y_1,x_2+y_2\rangle,\\ 
I_2(\textbf{x},\textbf{y};a,b)&=&b\langle x_1,x_2\rangle+a\langle y_1,y_2\rangle+\langle x_1,x_2\rangle \langle y_1,y_2\rangle.
\end{eqnarray}
%\endnumparts
In fact, both vectors of the inner product in $I_1$ are invariants. However, as in the vector generalisation of the Adler-Yamilov map, the invariants are not enough to claim Liouville integrability.

The Liouville integrability of the vector generalisations of the YB maps we consider in this section is an open problem.

\section*{Acknowledgements}
We would like to thank A. Veselov, T. Kouloukas, V. Papageorgiou, A. Dzhamay and G. Grahovski for the discussions and their comments, P. Xenitidis for helping to improve the text and special thanks to D. Tsoubelis for making available his computing facilitites. A.V.M. would like to acknowledge support from EPSRC (EP/I038675/1). S.K.R. would like to acknowledge William Right Smith scholarship and John E. Crowther scholarship.

\end{document}